\input iopppt
\pptstyle
\title{A comment on a paper by Carot et al.}
\author{Alan Barnes}

\address{Computer Science Group, School of Engineering and Applied Science,
Aston University, Aston Triangle, Birmingham B4 7ET, UK
\footnote\dag{E-Mail: \ \ {\tt barnesa{\bf @}aston.ac.uk}
\ \ \ Fax:\ \  {\rm +44 121 333 6215}}}

\jl{6}
\submitted

\date

\beginabstract
In a recent paper Carot et al.~considered carefully the definition of
cylindrical symmetry as a specialisation of the case of axial symmetry.
One of their propositions states that if there is a second
Killing vector, which together with the one generating the
axial symmetry, forms the basis of a two-dimensional Lie algebra, then
the two Killing vectors must commute, thus generating an Abelian group.
In this comment a similar result, valid under considerably weaker
assumptions, is recalled: any two-dimensional Lie transformation group
which contains a one-dimensional subgroup whose orbits are circles,
must be Abelian. 
The method used to prove this result is extended to apply to
three-dimensional Lie transformation groups. It is shown that the
existence of a 
one-dimensional subgroup with closed orbits restricts the Bianchi
type of the associated Lie algebra  to be  I (Abelian), II,
III, VII$_{q=0}$, VIII or IX. 
The relationship between the present approach and that of
the original paper is discussed.
\endabstract
\vfill\eject

\section{Introduction}
We begin by recalling a few relevant definitions and theorms.
Following Carter (1970) a spacetime $\cal M$ is said to have  
cyclical symmetry if and only if the metric is invariant under the
effective smooth action $SO(2) \times \cal M \rightarrow \cal M$ of the
one-parameter cyclic group $SO(2)$.
If, in a cyclically symmetric spacetime, the set of fixed points of
the isometry is non-empty then the set of fixed points is an
autoparallel time-like 
two-surface (Mars and Senovilla, 1993) and the spacetime is said to be
axially symmetric. Mars and Senovilla also show that the Killing
vector is spacelike in the neighbourhood of the axis and satisfies the
elementary flatness condition on the axis.

In a recent paper Carot, Senovilla and Vera (1999) consider in detail
the definition of cylindrical symmetry;  this paper will be referred to
below as CSV.  
Proposition 2 of CSV states that in  an axially
symmetric spacetime, if there is 
another Killing vector which, with the axial Killing vector generates
a two-dimensional isometry group, then the two Killing vectors commute.
Thus the isometry group is Abelian.  A similar result for stationary
axisymmetric spacetimes was proved by Carter (1970).
Both proofs rely heavily on the existence of an axis where the axial
Killing vector vanishes.  
CSV then use their result as the basis 
of a straightforward definition of a cylindrical symmetry without the
additional assumptions that are often made namely: that the group is
Abelian and that an elementary flatness condition holds at the
axis. Thus a spacetime is cylindrically symmetric if
it admits a two-dimensional isometry group with spacelike orbits (i.e. a
$G_2$ acting on $S_2$) containing an axial symmetry.  

Although the assumption of the existence of an axis is reasonable in
many circumstances, there are physically realistic situations where
an axis may not exist:  the axis may be singular and so not part of the
manifold, or the topology of the manifold may be such that no axis
exists. An example of the latter situation is a torus embedded in a
3-dimensional flat space; the torus is cyclically symmetric but the
Killing vector field generating the rotation does not vanish anywhere
on the surface. It is interesting therefore to consider whether a
result of the same ilk as Proposition 2 of CSV is valid for
cyclically symmetric spacetimes.

\section{Cyclically symmetric manifolds admitting a $G_2$}
Suppose $\cal M$ is cyclically symmetric and suppose $X_0$ is the
Killing vector field generating the isometry.   We restrict attention to the
open submanifold $\cal N$ of $\cal M$ on which $X_0$ is non-zero.
The orbit each point of $\cal N$  under the cyclic symmetry is a
circle and these circles are the integral curves of the Killing vector
field $X_0$. 
Let $\phi$ be a circular coordinate running from $0$ to $2\pi$ which
labels the elements of $SO(2)$ in the usual way.  Then we can
introduce a system of coordinates 
$x^i$ with $i = 1 \ldots n$ and $x^1 = \phi$  adapted to $X_0$ such that
$X_0 = \partial_\phi$.  These coordinates are determined only up to
transformations of the form
$$\tilde \phi = \phi + f(x^\nu) \qquad {\tilde x}^\mu = g^\mu(x^\nu)$$
where $f$ and $g^\mu$ are smooth functions and where Greek
indices take values in the range $2 \ldots n$.

Suppose now that the isometry group of $\cal M$ contains a
two-dimensional subgroup $G_2$ 
and that the Killing vector field $X_1$ together with $X_0$ forms a
basis of the associated Lie algebra with commutation
relations $[X_0\ X_1] = a X_0 + b X_1$, say, where $a$ and $b$ are
constants.  In an adapted coordinate system the commutation relations
reduce to 
$${\partial X_1^\mu \over \partial \phi} = b X_1^\mu \qquad 
{\partial X_1^1 \over \partial \phi} = a + b X_1^1 $$
A straightforward integration gives
$$\eqalign{
X_1^\mu = B^\mu(x^\nu)\e^{b\phi} \qquad  & X_1^1 = A(x^\nu)\e^{b\phi} - a/b 
\qquad  {\rm for\ } b \ne 0\cr
X_1^\mu = B^\mu(x^\nu) \qquad & X_1^1 = A(x^\nu) + a\phi \qquad  {\rm for\ } b = 0}$$
where $A$ and $B^\mu$ are arbitrary functions of integration.  The
solutions of these equations must be periodic in $\phi$ with period
$2\pi$ if $X_1$ is to be single-valued on $\cal N$.  Thus both $a$ and
$b$ must vanish and the subgroup $G_2$ must be Abelian.

Note that the dimensionality of the manifold, the existence of a
metric and the fact that transformation group is an isometry group are
not used in the proof; thus it has been proved that any
two-dimensional Lie transformation 
group which acts on an $n$-dimensional manifold $\cal M$ and 
which contains a one-dimensional subgroup acting cyclically on $\cal M$
must be Abelian.  This remarkably simple and general result is not
new\footnote\ddag{The author remembers seeing this result presented as
a footnote in a paper around 30 years ago, but is unable to locate the
precise reference.}, but is perhaps not widely known.

\section{Cyclically symmetric manifolds admitting a $G_3$}
Suppose now that the manifold $\cal M$ admits  a
three-dimensional Lie group $G_3$ of transformations which contains a
one-dimensional subgroup, generated by
the vector field $X_0$, acting cyclically on $\cal M$. 
Let $X_1$ and $X_2$ be  vector fields on $\cal M$ which together with
$X_0$ form a basis of the associated Lie algebra. 
Two cases arise: either the Lie algebra admits a two-dimensional
subalgebra containing $X_0$ or there is no such subalgebra.

In the former case, by the result of the previous section the
subalgebra is Abelian. We assume without loss of generality that $X_0$
and $X_1$ form a basis of this subalgebra. Hence the commutation
relations can be written in the form
$$[X_0\ X_1] = 0 \qquad [X_0\ X_2] = a X_0 + b X_1 + c X_2$$
where $a$, $b$ and $c$ are constants.  In terms of a
coordinate system adapted to $X_0$ these commutators reduce to the
differential equations
$${\partial X_1^\i \over \partial \phi} = 0 \qquad 
{\partial X_2^i \over \partial \phi} = a\delta^i_1 + b X_1^i + c X_2^i$$
A straightforward argument similar to that in the previous section
shows that if the solutions are to be 
periodic in $\phi$,  we must have  $a=b=c=0$. Thus $X_0$ commutes with
the other two basis vectors $X_1$ and $X_2$.
The remaining commutator takes the form
$$[X_1\ X_2] = d X_0 + e X_1 + fX_2$$
where $d$, $e$ and $f$ are constants. Three algebraically distinct
cases arise. If $d=e=f=0$ the group is Abelian (Bianchi type I). 
If $e=f=0$ but $d \ne 0$, after a rescaling of $X_1$ and/or $X_2$, we
may set $d=1$. The commutation relations now are
$$[X_0\ X_1] = 0 \qquad [X_0\ X_2] = 0 \qquad [X_1\ X_2] = X_0$$
which is the canonical form for an algebra of Bianchi type II (see for
example Petrov, 1969).
If $e$ and $f$ are not both zero, by a change of basis which leaves
$X_0$ fixed, we can set $d=e=0$ and $f=1$.
Thus  the commutators are
$$[X_0\ X_1] = 0 \qquad [X_0\ X_2] = 0 \qquad [X_1\ X_2] = X_2$$
which is (essentially) the canonical form for an algebra of Bianchi
type III.

If the Lie algebra has no two-dimensional subalgebra containing $X_0$
then, without loss of generality, we may choose the basis vectors
$X_1$ and $X_2$ such that the commutation relations become
$$[X_0\ X_1] = X_2 \qquad [X_0\ X_2] = a X_0 + b X_1 + c X_2$$
where $a$, $b$ and $c$ are constants.  
 In terms of a coordinate system adapted to $X_0$ these reduce to 
$${\partial X_1^\i \over \partial \phi} = X_2^i \qquad 
{\partial X_2^i \over \partial \phi} = a\delta^i_1 + b X_1^i + c X_2^i$$
From these equations it is easy to deduce that
$${\partial^2 X_2^i \over \partial \phi^2} 
   =  b X_2^i + c {\partial X_2^i \over \partial \phi}$$
Thus for solutions periodic in $\phi$ with period $2\pi$, we must have
$c=0$ and $b=-n^2$ for some positive integer $n$. By a redefinition of
the basis vector
$\tilde X_1 = X_1 + a/b X_0$, we can also set $a=0$ so that the
commutation relations become
$$[X_0\ X_1] = X_2 \qquad [X_0\ X_2] = -n^2 X_1 
\qquad [X_1\ X_2] = d X_0 + e X_1 + fX_2$$
where $d$, $e$ and $f$ are constants.  In a coordinate system adapted
to $X_0$ the basis vectors are given by
$$\eqalign{
X_0^i & = \delta^i_\phi \cr
X_1^i & = 1/n(A^i(x^\nu) \sin n\phi -B^i(x^\nu) \cos n\phi) \cr
X_2^i &= A^i(x^\nu) \cos n\phi + B^i(x^\nu) \sin n\phi}$$
where $A^i$ and $B^i$ are functions of integration.

The commutators must, of course, satisfy
the Jacobi identity
$$[X_0\ [X_1\ X_2]] + [X_1\ [X_2\ X_0]] + [X_2\ [X_0\ X_1]] = 0$$
which implies that
$$[X_0\ \ (d X_0 +e X_1 +f X_2)] = e X_2 -n^2 f X_1 = 0$$
Thus $e = f = 0$.  Three algebraically distinct cases arise.
If $d=0$ the group is of Bianchi type VII$_{q=0}$ which is, of course,
the isometry group of the Euclidean plane. Note that the standard form
of the commutation relations for Bianchi type VII$_{q=0}$ is obtained by means
of the change of basis 
$$\tilde X_0 = 1/n X_0 \qquad \tilde X_1 = nX_1 \qquad \tilde X_2 = X_2$$
For $d<0$ and $d>0$ the Bianchi type is VIII and 
IX respectively; which are of course the symmetry groups of
Lobachevskii plane and two-sphere respectively.
The change of basis required to reduce the
commutation relations to the canonical form for Bianchi types VIII and IX is
$$\tilde X_0 = 1/n X_0 \qquad \tilde X_1 = n/\epsilon X_1 
\qquad \tilde X_2 = 1/\epsilon X_2$$
where $\epsilon = n\sqrt{|d|}$.  

The methods above can be applied to the case of four (and higher)
dimensional Lie groups containing a one-dimensional cyclic subgroup.
Again the allowed forms of the commutation relations are considerably
restricted. These results will be presented elsewhere.

\section{A correction}
Senovilla (1999) has pointed out an error in the statement of Theorem
3 of CSV.  At his request a corrected statement of the theorem is
presented here. 

Given a $G_{3}$ on $T_{3}$ group that contains an orthogonally
transitive Abelian $G_{2}$ subgroup generated by an axial
$\vec \xi$ (such that its set of fixed points is non-empty)
and $\vec \eta$, and a third integrable timelike
Killing vector field $\vec \zeta$, then it follows that:
if $G_{3}$ is the maximal isometry group, then it must be Abelian;
if $G_{3}$ is non-Abelian, then it is (locally) contained in a
$G_{4}$ on $T_{3}$, and $\vec \xi$ and $\vec \zeta$ generate an
orthogonally transitive subgroup $G_{2}$ on $T_{2}$.

The proof follows easily from the remarks immediately preceding
the statement of the theorem in the original paper.

\section{Summary}
Low-dimensional Lie transformation groups which act on an $n$-dimensional
manifold $\cal M$ and which contain a one-dimensional subgroup acting
cyclically on $\cal M$ have been considered. For the two-dimensional
case the group must be Abelian and for the three-dimensional case the
Bianchi type of the group is restricted: types IV, V and VI cannot
occur and  only the subclass VII$_{q=0}$ of type VII is allowed. As
far as the author is aware the results for the three-dimensional case
are new.   

At first sight the restrictions in the three-dimensional case 
may not seem too severe: only three of the nine Bianchi types are
excluded whilst a fourth type is restricted. However, it should be
pointed out that 
Bianchi types VI and VII both involve an arbitrary parameter and so
each contain an infinite number of algebraically distinct types. Those
in type VI are excluded completely and for type VII only a
single case survives. Moreover for most of the types which {\it
are\/} permitted 
the structure of the Lie algebra is `aligned' in some way relative to $X_0$.
For example, for Bianchi type II, $X_0$ is also the generator of the
first derived subalgebra and for Bianchi types II and III, $X_0$ lies in
the centre of the algebra.

At first sight too, the result presented in section 2 seems considerably
stronger than Proposition 2 of CSV; however this is only
partially true. Certainly the results presented do not require the
existence of an axis and they apply to any Lie
transformation group.  So {\it a fortiori\/} the results apply to isometries,
conformal motions, affine and projectives collineations etc.~whereas
CSV's results apply only to isometries (although some of the results,
at least, may be extended to conformal motions). However the results
presented here require that the transformations are defined everywhere
on the manifold and so they say nothing about the structure of groups of
{\it local\/} transformations.  CSV's results apply to local
isometries in the neighbourhood of the axis and so in this sense are
stronger than the results presented above.  Of
course, if we assume that the isometries are globally well-defined
and that an axis of fixed points exists, it may well be possible to
derive stronger restrictions on the allowed group structure  using
a synthesis of the two methods. This is currently under investigation. 

\references 
\refjl{Carot J, Senovilla J M M and Vera R}{\CQG}{16} {3025--34}
\refjl{Carter B 1970}{Commun. math. Phys.}{17}{233--8}
\refjl{Mars M and Senovilla  J M M 1993}{\CQG}{10}{1633--47}
\refbk{Petrov A Z 1969}{Einstein Spaces}{Pergamon Press, Oxford, p.~63}
\refbk{Senovilla J M M 1999}{Private Communication}{}
\bye